\documentclass[doublecol]{epl2} 

\usepackage{amsmath}
\usepackage{amssymb}
\usepackage[normalem]{ulem}

\title{Unusual giant magnetostriction in the ferrimagnet Gd$_{2/3}$Ca$_{1/3}$MnO$_3$}
\shorttitle{Unusual giant magnetostriction in the ferrimagnet Gd$_{2/3}$Ca$_{1/3}$MnO$_3$} 

\author{V. F. Correa\inst{1} \and N. Haberkorn\inst{1} \and G. Nieva\inst{1} \and D. J. Garc\'ia\inst{1} \and B. Alascio\inst{1}}
\shortauthor{V. F. Correa \etal}

\institute{                    
  \inst{1} Centro At\'omico Bariloche (CNEA) and Instituto Balseiro (U. N. Cuyo), 8400 Bariloche, R\'io Negro, Argentina}

\abstract{
We report an unusual giant linear magnetostrictive effect in the ferrimagnet Gd$_{2/3}$Ca$_{1/3}$MnO$_3$ ($T_{c}\!\approx$80 K). 
Remarkably, the magnetostriction, negative at high temperature ($T \approx T_{c}$), becomes positive below 15 K when the magnetization of the Gd sublattice overcomes the magnetization of the Mn sublattice.
A rather simple model where the magnetic energy competes against the elastic energy gives a good account of the observed results and confirms that Gd plays a crucial role in this unusual observation.
Unlike previous works in manganites where only striction associated with 3$d$ Mn orbitals is considered, our results show that the lanthanide 4$f$ orbitals related striction can be very important too and it cannot be disregarded.}

\begin{document}

\maketitle

\section{Introduction}

Manganites are perovskites mostly known for their spectacular colossal magnetoresistance (CMR): the electrical resistivity can change several orders of magnitude under a moderate applied magnetic field $B$ \cite{Jin}. 
They also show another impressive property called giant linear magnetostriction (MS): sample dimensions are strongly affected by a magnetic field, either external or molecular \cite{Ibarra}. The effect is comparable in magnitude ($\Delta L/L \!\geq$10$^{-3}$ at several Tesla) to the highest MS values ever reported.
Both CMR and MS are particularly large around the $Mn$-ions ferromagnetic ordering temperature \cite{Kimura}. Associated with this order, and depending on the doping level, manganites can display a metal-insulator (MI) transition, too.
In this way, manganites offer a unique testing ground to study the interplay between electronic, spin and lattice degrees of freedom. 

As expected, structural distortion of the plain perovskite structure strongly affects the magnetic and electronic properties of manganites. This is usually parametrized by the so called tolerance factor $t$, which quantizes the mismatch between the size of the different ions in the formula. This mismatch primarily influences the exchange interaction between Mn ions altering both the length and the angle of the Mn-O-Mn bond.
An `\textit{universal}' temperature $T$ versus $t = (d_{R/A-O}) / \sqrt{2}(d_{Mn-O})$ phase diagram has long been reported \cite{Hwang} for the hole doped R$_{2/3}$A$_{1/3}$MnO$_3$ manganites ($R$ is a lanthanide and $A$ is an alkaline-earth element). Slightly distorted structures ($t\!\sim$1) show the insulating-paramagnet (PMI) to metallic-ferromagnet (FMM) transition. However, the metallic state disappears at higher distortions ($t$$\lesssim$0.91) even though a transition to a insulating-ferromagnet (FMI) is observed.  

An estimated value of $t\!\approx$0.89 places Gd$_{2/3}$Ca$_{1/3}$MnO$_3$ well inside the insulating regime. Indeed, no MI transition is observed down to 5 K with the resistivity $\rho$ showing a characteristic semiconducting behavior in the whole temperature range \cite{Snyder,Hueso}. Nevertheless, magnetic properties are quite more interesting. Mn magnetic moments start ordering ferromagnetically around $T_c\!\sim$80 K. The Gd moments react to the internal field created by the Mn ferromagnetic sublattice gradually aligning in the opposite direction. The two sublattices compete each other giving rise first to a maximum in the magnetization around 50 K and finally to a full compensation at $T_{comp} \!\sim$15 K where the magnetization vanishes. At lower temperature, the Gd magnetic moment overcomes the Mn moment. The overall temperature dependence of the magnetization corresponds then to a ferrimagnet created by the two opposite Mn and Gd sublattices \cite{Snyder,Pena,Zukrowski,Haberkorn}.  

In this work we study the magnetostructural properties of Gd$_{2/3}$Ca$_{1/3}$MnO$_3$. The rather complex magnetic structure clearly couples to the atomic lattice giving rise to a giant linear magnetostrictive effect \cite{Correa}. Remarkably, the negative field dependence of the MS observed at high temperature changes its sign and becomes positive when $T<T_{comp}$. 
We use a 4-site mean field approximation to model the experimental data. It demonstrates that the competition between Gd-Gd and Gd-Mn spin correlations is responsible of the sign change in the MS. 
This finding shows that the usually underestimated MS associated with the lanthanide 4$f$ orbitals in manganites can be comparable to the usual giant striction given by the re-orientation of the Mn 3$d$ orbitals.

\section{Experimental details}

Pure single crystalline samples of Gd$_{2/3}$Ca$_{1/3}$MnO$_3$ were grown by the floating zone technique. Crystal quality and composition have been checked through XRD and EDS scans. 
A capacitive technique was used in the dilation experiments. The high resolution ($\leq$1 \AA) dilatometer \cite{Schmiedeshoff} is placed in a evacuated environment with a low pressure ($P<$10$^{-1}$ torr) of exchange He$^4$ gas.
Magnetic field is applied along the [020] direction of the orthorhombic $Pnma$ crystalline structure  ($a$= 5.39\AA, $b$= 5.56\AA and $c$= 7.5\AA) in all the experiments. Dilation experiments are always performed in a longitudinal configuration with $B \parallel L \parallel [020]$. Several samples of different sizes have been measured with a perfect agreement between them. Sample length $L$ is typically about 200 $\mu$m.

\section{Results and discussion}

\begin{figure}
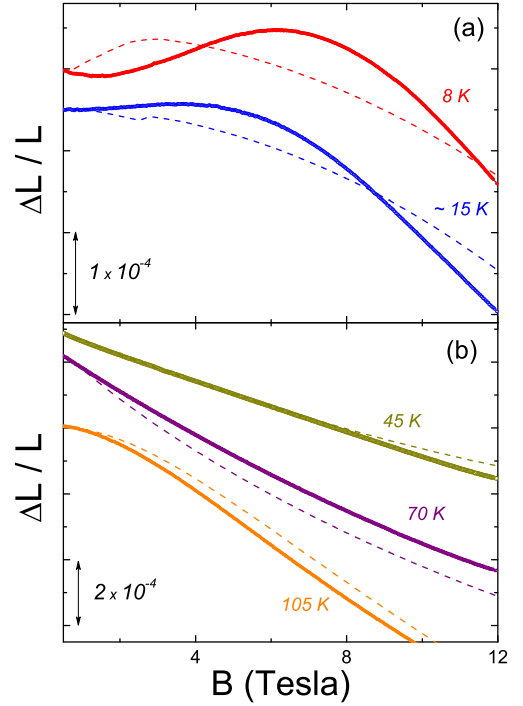

\onefigure{fig1.eps}
\caption{(color online) Experimental (solid) and calculated (dashed) 
magnetostriction. Upper (lower) panel shows results in the low (high) temperature range. Curves are vertically shifted.}
\label{fig1}
\end{figure}

Representative isothermal linear magnetostriction results after a zero field cooling procedure are shown in Fig. \ref{fig1} (solid lines). The effect is giant with no evidence of saturation up to $B$ = 12 T (the highest applied field), reaching a maximum value around $T_c\!\sim$80 K ($\Delta L/L \!\approx$10$^{-3}$).  Two very distinctive regimes are found: 

(i) above $T_{comp} \sim$15 K the field dependence of $L$ is negative and monotonic. Hysteresis and relaxation effects are important, mainly in the range 40 K$\lesssim T \lesssim\! T_{c}$, 

(ii) below $T_{comp}$ magnetostriction becomes positive at low fields $B \leq$ 7 T (the initial negative slope at $B <$ 1 T is associated to magnetic domains and is absent if the experiment is performed after a field cooling procedure). At higher fields it turns negative again resulting in an overall non-monotonic field dependence of the MS. On the other hand, around $T_{comp}$ (where the magnetization almost vanishes), the magnetostriction is negligible below $B\! \sim$4 T. 

$T_{comp}$ marks the onset of the Gd magnetic ordering which dominates the low temperature regime while Mn moments prevail in the high temperature regime. In this sense, this unusual magnetostriction strongly points toward the interplay of the different magnetic interactions: Mn-Mn, Mn-Gd and Gd-Gd.
We use a simple model to verify this hypothesis where the three different interactions are introduced in the Hamiltonian via Heisenberg-like terms.

We consider an homogeneous network of Gd ions with one Gd ion for each Mn, ignoring the random nature of their localizations.
As there are 2/3 Gd ions for each Mn, we rescale the Gd effective magnetic moment to J$=2/3\times7/2=7/3\sim 5/2$.
We choose the $5/2$ value for the Gd spin to retain the quantum nature of the spin without rescaling of the $g$-factor ($g_{Gd}=2$).
Manganese ions appears in a mixture of $1/3$ of S=3/2 and $2/3$ of S=2. 
As in both cases the orbital magnetic moment is quenched we take $g_{Mn}=2$.
To keep the experimental zero temperature net magnetic moment of 1 $\mu_B$ (perfect ferrimagnetic ordering given by the two sublattices \cite{Snyder}) we take S=2 for the Mn effective magnetic moment.

Based on these considerations we write a Hamiltonian with a ferromagnetic (coupling $K_{Mn-Mn}$) network of manganese spins $S$ (S=2) antiferromagnetically coupled ($K_{Mn-Gd}$) to a network of ferromagnetic ($K_{Gd-Gd}$) gadolinium spins $J$ (J=5/2).
The magnetic interaction is given by the Hamiltonian:
\begin{eqnarray}
H_{m} = K_{Mn-Mn} \sum_{\left\langle i,j\right\rangle} S_i \cdot S_j + K_{Mn-Gd} \sum_{i} S_i \cdot J_i \nonumber  \\ 
+ K_{Gd-Gd} \sum_{\left\langle i,j\right\rangle} J_i \cdot J_j + g \mu_B \vec{B}\cdot \sum_{i} (S_i + J_i) \label{magham}
\label{Hm}
\end{eqnarray}

The smaller values of the effective spins allow us also to use a 4-sites (2 Mn and 2 Gd) cluster in the Constant Coupling approximation (see Appendix) \cite{CC1,CC2,Davies}.
We consider six neighbours ($z=6$). In the Constant Coupling approximation the interactions are isotropic, meaning that they represent averaged interactions. 

Following early works \cite{Davies,Doerr,Zapf}, we consider that the exchange parameters are strain dependent. If the lattice is under some small distortion $\delta_L$ all the coupling parameters change accordingly:
\begin{eqnarray}
K_{Mn-Mn} & = & K^0_{Mn-Mn}+\alpha \delta_L \nonumber \\
K_{Mn-Gd} & = & K^0_{Mn-Gd}+\beta \delta_L \\
K_{Gd-Gd} & = & K^0_{Gd-Gd}+\gamma \delta_L \nonumber
\end{eqnarray}

From fittings to magnetization experiments, we obtain: $K^0_{Mn-Mn} = −$-9 K,  $K^0_{Mn-Gd} = $ 8 K and $K^0_{Gd-Gd} = $ 0 K.
Mn-Mn coupling $K^0_{Mn-Mn}$ is ferromagnetic with $T_c$ close to the experimental value of $\sim 80$ K;  $K^0_{Mn-Gd}$ is antiferromagnetic and gives $T_{comp}\sim 15$ K together with an effective null coupling between Gd ions (also expected due to the dipolar origin of those interactions). 
Both, our experimental and calculated magnetization results are similar to those reported by Snyder et al. \cite{Snyder}. 

The presence of a distortion also increases the elastic energy
\begin{equation}
E_{e} = 1/2 C \delta_L^2
\end{equation}

For a state $\vert G \rangle$, the total energy $E_{m}+E_{e}$ is minimized for 
\begin{equation}
\delta_L (B) = - \frac{1}{C} \langle G \vert \alpha \sum S_i \cdot S_j + \beta \sum S_i \cdot J_i + \gamma \sum J_i \cdot J_j \vert G \rangle_{B}  \nonumber
\end{equation}

As we are interested in the length distortion respect to the $B=0$ case, we compute
{\setlength\arraycolsep{1pt}
\begin{eqnarray}
\frac{\Delta L}{L} &=& \delta_L (B)-\delta_L (0) \nonumber \quad \\
  &=& \Lambda_{\alpha} (\langle G \vert \sum_{\left\langle i,j\right\rangle} S_i \cdot S_j \vert G \rangle_{B=0}-\langle G \vert \sum_{\left\langle i,j\right\rangle} S_i \cdot S_j \vert G \rangle_{B}) \nonumber \quad \\
 &+&\, \Lambda_{\beta} (\langle G \vert \sum_{i} S_i \cdot J_i \vert G \rangle_{B=0}-\langle G \vert \sum_{i} S_i \cdot J_i \vert G \rangle_{B}) \quad \\
 &+&\, \Lambda_{\gamma} (\langle G \vert \sum_{\left\langle i,j\right\rangle} J_i \cdot J_j \vert G \rangle_{B=0}-\langle G \vert \sum_{\left\langle i,j\right\rangle} J_i \cdot S_j \vert G \rangle_{B}) \nonumber \quad
\end{eqnarray}}

\noindent where $\Lambda_{\chi} = \frac{\chi}{C}$ and $\chi=\left\{\alpha,\beta,\gamma\right\}$. $\langle \rangle_{B}$ denotes the thermal expectation value.  

\begin{figure}[h]
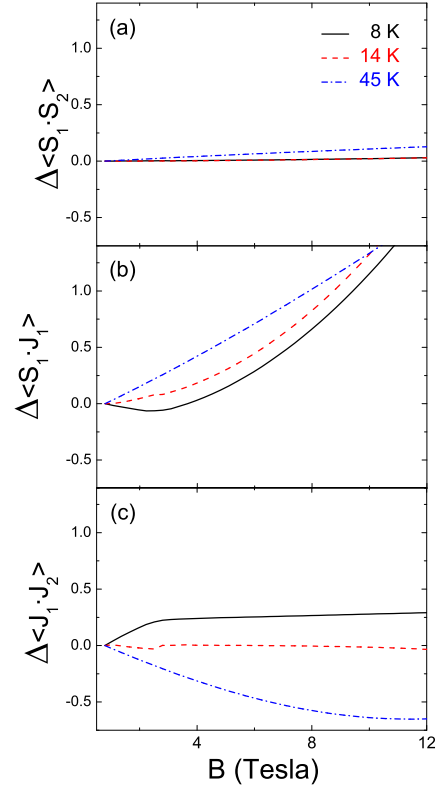

\onefigure{fig2.eps}
\caption[]{(color online) Computed correlations change respect to the zero field situation 
for 8 K (below $T_{comp}$), 14 K (close to $T_{comp}$) and 45 K (above 
$T_{comp}$). (a) Mn-Mn correlations, (b)Mn-Gd correlations and (c) Gd-Gd correlations.}
\label{fig2}
\end{figure}

The correlations are computed using the 4-site Constant Coupling approximation and correspondingly now $i,j=1,2$. Figure \ref{fig2} shows the computed spin correlators as a function of magnetic field ($\Delta \langle O \rangle = \langle O \rangle_{B} - \langle O \rangle_{B=0}$, where $O=S_1\cdot S_2,S_1\cdot J_1$ or $J_1\cdot J_2$) in the different temperature ranges: below, around and above $T_{comp}$. 

There are several issues to emphasize:

i) the field dependence of the Mn-Mn correlations ($\Delta \langle S_1 \cdot S_2 \rangle$) is monotonic and, except around $T_c$, it is also very small (Fig. \ref{fig2}(a)), reflecting the fact that Mn sublattice is almost fully polarized at low temperature.

ii) Mn-Gd correlations ($\Delta \langle S_1 \cdot J_1 \rangle$) show the largest effect below $T_c$, as seen in Fig.  \ref{fig2}(b). This is reasonable since the applied field tends to align both sublattices in the same direction gradually destroying the otherwise almost perfect low temperature ferrimagnet.
Its contribution to the low field total magnetostriction, however, is smaller than that due to Gd-Gd correlations by a factor $\frac{1}{4}$ approximately. At higher fields, on the other hand, $\Delta \langle S_1 \cdot J_2 \rangle$ becomes the larger contribution.   

iii) Gd-Gd correlations ($\Delta \langle J_1 \cdot J_2 \rangle$) show the most relevant and distinct behavior. While $\Delta \langle J_1 \cdot J_2 \rangle$ is negative for $T > T_{comp}$, it vanishes around $T_{comp}$ and becomes positive when $T < T_{comp}$. 
The physical origin of this sign change is related to the local field acting on Gd moments.
This local field is made up of the molecular field created by the Mn moment $B_{Mn}$ plus the external field $B$. 
Above $T_{comp}$, $B_{Mn}$ points opposite to $B$ so an increasing $B$ ($B<B_{Mn}$) results in a decreasing local field. 
This decreasing local field reduces the Gd moment and consequently the Gd-Gd correlations.
Below $T_{comp}$, on the other hand, $B_{Mn}$ points in the same direction of $B$, so an increasing $B$ ($B<B_{Mn}$) results in a increasing local field. This increasing local field raises Gd-Gd correlations.
This sign change in the Gd-Gd correlations is indeed responsible for the sign change in the magnetostriction observed at low temperature.

Computed MS curves are also shown in Figure \ref{fig1} ($\Lambda_{\alpha} = 12 \times 10^{-4}$, $\Lambda_{\beta} = 0.8 \times 10^{-4}$ and $\Lambda_{\gamma} = -1.6 \times 10^{-4}$). $\Lambda_{\alpha}$ is chosen as to get a good agreement at high temperature ($T \geq T_c$) where the only non-negligible correlator is $\Delta \langle S_1 \cdot S_2 \rangle$  while $\Lambda_{\gamma}$ is chosen as to get a positive striction at low temperature and field ($T < T_{comp}$, $B \leq 5$ T) where only $\Delta \langle J_1 \cdot J_2 \rangle$ is non-negligible. $\Lambda_{\beta}$ is then selected to get the best agreement in the whole temperature and field range. 

The model gives a good account for the non-monotonic $\Delta L / L$ below $T_{comp}$ (see curve at 8 K, Fig. \ref{fig1}(a)). It is a consequence of two opposite contributions: a negative Gd-Gd magnetostructural coupling ($\Lambda_{\gamma} < $ 0) and a positive Mn-Gd coupling ($\Lambda_{\beta} >$ 0). As stated previously, Mn-Mn correlations are almost saturated and they do not contribute to the MS in this low temperature range.
The coupling parameters ($K$´s) used (fixed by the fit of the magnetic properties) allows us to reproduce only qualitatively the field value of the MS maximum. 
The model also accounts for the extinction of this maximum at $T_{comp} \sim $ 15 K, where the striction becomes very small.

At higher $T$ ($T_{comp} \leq T \ll T_c$) the MS gets negative. As before, Mn-Mn correlations almost do not change but now both Gd-Mn and Gd-Gd correlation effects point in the same direction. 
In the intermediate and high temperature range ($T \gg T_{comp}$), where the Mn-Mn contribution is the more relevant one, the agreement is fairly good, except around $T_c$. Not only the magnitude of the magnetostriction is well 
accounted, also the curvature of the isotherms is very well reproduced (see Fig. \ref{fig1}(b)).
It is interesting to stress that even though magnetization is an increasing function of field in the whole temperature range, the magnetic correlations and so magnetostriction shows two very distinctive regimes: a monotonic magnetostriction at high temperature that becomes non-monotonic below $T_{comp}$.

In this isotropic model, $C$ can be estimated as $v B_T$, where $v$ is the volume of the perovskite unit cell and $B_T$ is the Bulk modulus. 
Mn-Mn parameter $\Lambda_{\alpha}$ is positive, and so is $\alpha$. Since the interaction is FM ($K_{Mn-Mn} < $ 0), that implies that $\left|K_{Mn-Mn}\right|$ increases as the lattice gets smaller. This is the expected behavior for exchange-like interactions. 
There are no available pressure effects on Gd$_{2/3}$Ca$_{1/3}$MnO$_3$ to compare with. Nevertheless, a rough estimate can be done by replacing Gd by another lanthanide, i.e. by chemical pressure. This substitution (keeping the composition at R$_{2/3}$Ca$_{1/3}$MnO$_3$) does not modify the Mn valence and so, its magnetic moment. So, a change in $T_c$ can in principle be associated with an inter-ion distance $d$ change.
In this isotropic approximation, $d = v^{1/3}$. Taking $B_T =$ 150 Gpa \cite{Srivastava}, and $v = 55.36 \AA^3$,\cite{Pena} we get $\alpha = 722$ K.
For Dy$_{2/3}$Ca$_{1/3}$MnO$_3$, $v_{Dy} \approx 55.10 \AA^3$ \cite{Pena2}. This results in a change of the exchange parameter $K_{Mn-Mn}$ given by $\Delta K = \alpha \frac{d_{Dy} - d_{Gd}}{d_{Dy}} \approx$ -2 K. This very rough estimate of a 20 percent increase in the exchange parameter ($K_{Mn-Mn} = $ -9 K) is of the same order of magnitude that the $T_c$ increase observed in Dy$_{2/3}$Ca$_{1/3}$MnO$_3$ \cite{Pena3}.

Gd-Gd parameter $\Lambda_{\gamma}$ is negative and it may be related to the dipolar (anisotropic) origin of these interactions: for zero distortion the net (average) interaction is zero but when the lattice shrinks AF interactions prevails.
Notably, the magnetostriction associated with Gd (low field positive striction below $T_{comp}$) is of the same order of magnitude than the observed striction in metallic gadolinium \cite{Davies}. 
On the other hand, Mn-Gd parameter $\Lambda_{\beta}$ is positive and it is much more difficult to understand since Mn-Gd interactions are antiferromagnetic: this effective interaction diminishes as the lattice shrinks.
This counter-intuitive value of $\Lambda_{\beta}$ could be related with the rotation of the oxygen tetrahedra that sourrounds Mn ions.

\section{Conclusions}
     
An unusual non-monotonic giant magnetostriction is observed in single crystals of the ferrimagnet Gd$_{2/3}$Ca$_{1/3}$MnO$_3$ at low temperature ($T < T_{comp} \!\sim$15 K) arising from the interplay between the Mn and Gd magnetic sublattices.
A simple mean field approximation where different magnetic interactions (Gd-Gd, Mn-Mn and Mn-Gd) compete among them and with the elastic energy gives a good account of the observed results.
Particularly, the change in the sign of the magnetostrictive effect at low temperature is driven by the competition between Mn-Gd and Gd-Gd magnetic correlations and it does not involve the Mn-Mn correlations.
Unlike previous works \cite{deTeresa} in manganites where only striction associated with $d$ orbitals is considered, our results show that $f$ orbitals related striction can be as important.

\section{Appendix}

\subsection{Constant Coupling Approximation \label{CF}}

The constant coupling (CC) approximation \cite{CC1,CC2,Davies} is an improvement over a classical mean field approximation.
It allows to consider correlations and to obtain a critical temperature closer to the exact result.
A classical mean field approximation replaces all the interactions of a site by an effective magnetic field made up of the external field and its neighbours magnetization. This neighbours magnetization is assumed to be the same than  that of the site.
In the constant coupling approximation two systems must give identical results for the magnetization.
If the original problem is in a network with $z$ neighbours per site, one system is made with a single site and $z$ ``effective'' neighbours, while the other is made up with a cluster of 2 sites and $z-1$ ``effective'' neighbours for each site.
In a classical mean field we have to search for an effective field proportional to the neighbours magnetization.
The CC approximation consists in searching for an effective field in both systems such that the same magnetization is obtained in the single site and the cluster.

\begin{figure}[h]
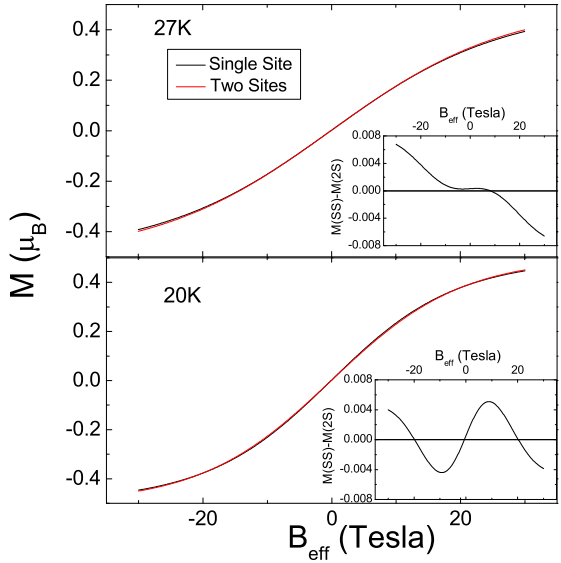

\onefigure{fig_ap.eps}
\caption{(color online) Magnetization as a function of the effective field for a single site and a two site cluster (S=1/2, $K=30K$, $z=6$). Upper panel: 27 K. Lower panel: 20 K. Insets show the magnetization difference. The classical mean field transition temperature is $T_{C,MF}= 2/3 z S (S+1) K=90$ K.} 
\label{fig_ap}
\end{figure}

To illustrate the procedure we use a simple spin 1/2 ferromagnet.
For a single site the Hamiltonian is written as
\begin{equation}
H_{ss} =   g \mu_B (z \vec{B}_{eff} + \vec{B}) \cdot S  \nonumber
\label{Hss}
\end{equation}

\noindent where $B_{eff}$ is an effective field. In a classical mean field approximation $B_{eff}=K_{Mn-Mn} \langle S \rangle$ and a self-consistent $\langle S \rangle$ is looked for.
For a two sites cluster the Hamiltonian is
\begin{equation}
H_{2s} = K S_1 \cdot S_2 + \, g \mu_B \left[ (z-1) \vec{B}_{eff} + \vec{B} \right] \cdot \left[S_1+S_2\right]    \nonumber
\label{H2s}
\end{equation}

In the upper panel of Fig. \ref{fig_ap} we show the magnetization at high temperature (27 K, just above the transition) of both a single site and a cluster of two sites as a function of $B_{eff}$ with an applied external field $B=0.1$ T.
These magnetizations are zero for $(z-1)B_{eff}=-B$ (or $z B_{eff}=-B$ for a single site) and show the expected paramagnetic-like behaviour for small systems.
At temperatures above the transition temperature both magnetizations agree for a single field which is the searched effective field (around 7 T for this temperature; the corresponding magnetization is less than $0.1\mu_B$).
As the temperatures lowers, this solution moves toward higher fields with a corresponding larger magnetization.
Below the transition temperature, two new solutions (higher in energy) appears, just as in a regular mean field approximation (lower panel of Fig. \ref{fig_ap}).

For the ferrimagnet Gd$_{2/3}$Ca$_{1/3}$MnO$_3$ we take as the ``single'' site an unit made up of an effective manganese (S=2, $g=2$) and an effective Gd (J=5/2, $g=2$) ion
\begin{eqnarray}
H_{ss} &=&  g \mu_B (z \vec{B}_{eff,Mn} + \vec{B}) \cdot S  \nonumber \\
& & +  \,  g \mu_B (z \vec{B}_{eff,Gd} + \vec{B}) \cdot J  \nonumber \\
& & +  \, K_{Gd-Mn} S \cdot J  \nonumber
\label{HssFM}
\end{eqnarray}

The cluster is made with two Mn and two Gd ions and we take $z=6$.
Each Mn (Gd) ion interacts with the other and with an external field made by the $z-1$ remaining neighbours.
In each site, there is a Gd-Mn interaction.
The Hamiltonian is 
\begin{eqnarray}
H_{2s} &=& K_{Mn-Mn} S_1 \cdot S_2 + K_{Gd-Gd} J_1 \cdot J_2 \nonumber \\
			& & + \, K_{Gd-Mn} (S_1 \cdot J_1 + S_2 \cdot J_2) \nonumber  \\
			& & + \, g \mu_B \left[ (z-1) \vec{B}_{eff,Mn} + \vec{B} \right] \cdot \left[S_1+S_2\right]  \nonumber \\ 
	    & & + \, g \mu_B \left[ (z-1) \vec{B}_{eff,Gd} + \vec{B} \right] \cdot \left[J_1+J_2\right]  \nonumber 
\label{H2sFM}
\end{eqnarray}

When the effective field is found, correlations between the different ions can be computed in the 2-site (four ions) cluster.

\acknowledgments
We thank M. T. Causa, L. Manuel, C. Balseiro and A. Aligia for fruitful discussions.
The authors are member of CONICET, Argentina. Work partially supported by ANPCyT PICT05-32900, PICT07-00812, PICT07-00819, PICT08-1043 and SeCTyP-UNCuyo 06/C326.

\end{document}